\documentclass[12pt]{article}

\usepackage[dvips]{epsfig,graphicx, color}

\usepackage{lscape}
\usepackage{times}
\usepackage{amsmath, amssymb}

\newenvironment{sciabstract}{%
\begin{quote} \bf}
{\end{quote}}

\newcounter{lastnote}

\title{A quantitative study of the benefits of co-regulation using the \textit{spoIIA} operon as an example
}

\author{Dagmar Iber \\
{\small Mathematical Institute, Centre for Mathematical Biology} \\
{\small 24-29 St. Giles, Oxford OX1 3LB, UK} \\
{\small Tel.: +44-1865-283875, Fax: +44-1865-273583,
iber@maths.ox.ac.uk}
              }
\date{}

\begin{document}
\maketitle
%\baselineskip24pt
% Include your paper's title here

\newpage
% Place your abstract within the special {sciabstract} environment.
\begin{sciabstract}
%\section*{Summary}
The distribution of most genes is not random, and functionally
linked genes are often found in clusters. Several theories have been
put forward to explain the emergence and persistence of operons in
bacteria \cite{Lawrence97}. Careful analysis of genomic data favours
the co-regulation model \cite{Pal04,Price05}, where gene
organization into operons is driven by the benefits of coordinated
gene expression and regulation. Direct evidence that co-expression
increases the individual's fitness enough to ensure operon formation
and maintenance is, however, still lacking. Here, a previously
described quantitative model of the network that controls the
transcription factor $\sigma^F$ during sporulation in
\textit{Bacillus subtilis} \cite{Iber_Spo_nature} is employed to
quantify the benefits arising from both organisation of the
sporulation genes into the \textit{spoIIA} operon and from
translational coupling. The analysis shows that operon organization,
together with translational coupling, is important because of the
inherent stochastic nature of gene expression which skews the ratios
between protein concentrations in the absence of co-regulation. The
predicted impact of different forms of gene regulation on fitness
and survival agrees quantitatively with published sporulation
efficiencies.
\end{sciabstract}

\textbf{Running Title} A quantitative study of the benefits of
co-regulation


\textbf{Subject category} Chromatin $\&$ Transcription$/$Metabolic
and regulatory networks

\textbf{Characters} 19650\\

\textbf{standfirst text}\\
The benefits of co-regulated gene expression have been suggested to
drive operon emergence and persistence but direct evidence that
co-expression increases an individual's fitness is lacking. Here, a
previously described quantitative model of the $\sigma^F$ signaling
network is employed to show that the inherent noise in gene
expression can be sufficiently harmful that co-regulated
expression can substantial increase survival chances.\\

\textbf{Main findings of the study}\\
\begin{itemize}
\item the study provides further support for the co-regulation model
for operon formation
\item the study reveals that small variations in gene expression, as
arise from the inherent stochasticity of biological processes, can
be harmful, and that co-regulation of the expression of interacting
proteins by organization of the genes into operons can substantially
increase survival chances
\item the quantification of the impact of co-regulation on an individual's
fitness is possible for the first time because of the detailed
mathematical model that we have developed recently for the genes
encoded in the spoIIA operon.

\end{itemize}

\newpage

\section{Introduction}
The available genome sequences demonstrate that many genes are
clustered on chromosomes according to their function. Genes in
bacteria are clustered but can also be organized into operons such
that the expression of a group of genes is regulated by the same
genetic control element. When operons were first discovered it was
assumed that the benefit of co-transcription led to operon assembly
\cite{Jacob61}. Other models have since been proposed, and these
belong to one of three classes, the natal model, the Fisher model,
or the selfish operon model \cite{Lawrence97}. According to the
natal model, clustering of genes is the consequence of gene
duplication. However, since operons comprise genes that belong to
very distant families and the majority of paralogs do not cluster,
this model is insufficient to explain operon origin
\cite{Lawrence97,Dandekar98}. A recast of the Fisher model, adapted
to prokaryotes, proposes that clustering of genes reduces the
likelihood that co-adapted genes become separated by recombination.
However, this does not explain how operons can emerge, as
recombination is as likely to generate clusters as to disrupt them.
According to the selfish operon model, operons facilitate the
horizontal transfer of functionally related genes \cite{Lawrence96}.
The physical proximity of genes thus does not necessarily provide a
selective advantage to the individual organism but rather to the
gene cluster itself, because it can be efficiently transmitted both
horizontally as well as vertically. Recent studies have, however,
failed to observe the gene cluster pattern predicted by the model,
and this strongly suggests that the selfish operon model does not
explain the emergence and persistence of operons
\cite{Pal04,Price05}. So what drives operon assembly?

The idea that co-transcription of genes provides a selective
advantage to the individual organism has never been contradicted. It
has been questioned only because it remains unclear whether the
benefits of co-transcription could be strong enough to drive the
assembly of operons by rare recombination events
\cite{Lawrence96,Lawrence97}. A genotype that confers higher fitness
will dominate in a population with bounded total population size
only if selection acts on a timescale that is substantially shorter
than the timescale on which recombination and mutation events could
negate the benefits.

There are a number of potential selective advantages given by
co-transcription. In the case of operons that code for multi-protein
complexes, co-transcription enables co-translational folding
\cite{Dandekar98}, it limits the half-life of toxic monomers
\cite{Pal04}, and it reduces stochastic differences in gene
expression \cite{Swain04}. Operons that do not code for interacting
proteins may be advantageous because of the co-regulation of protein
expression \cite{Price05}. Many examples of this class of operons
are associated with metabolic operons \cite{Lawrence96} where
co-regulated expression is likely to optimize the flux and to
facilitate the regulation of functions, especially if these are
required only under certain environmental conditions, or if complex
regulatory structures are employed \cite{Price05}.

Evidence in favour of any of these proposed driving forces has so
far largely been obtained from comparative genomics. Here we use a
previously derived quantitative model for the network that controls
the transcription factor $\sigma^F$ during sporulation in
\textit{Bacillus subtilis} \cite{Iber_Spo_nature} to quantify the
benefits of co-expression. Spore formation in \textit{Bacillus
subtilis} is a response to nutrient deprivation at high cell density
and involves asymmetric septation and compartment-specific
initiation of gene expression \cite{Hilbert04}. The different gene
programs in the larger mother cell and the smaller prespore are both
directed by the transcription factor $\sigma^F$ which, although only
active in the smaller prespore, affects the transcriptional programs
across the septum also in the mother cell, a phenomenon that is
referred to as criss-cross regulation \cite{Losick92}. Successful
sporulation therefore requires the rapid septation-dependent and
prespore-specific activation of $\sigma^F$. $\sigma^F$ is kept
inactive by binding to SpoIIAB and is released upon binding of
SpoIIAA (Fig. 1). SpoIIAA is phosphorylated by SpoIIAB \cite{Min93}
and reactivated by the serine phosphatase SpoIIE \cite{Duncan95}.
The balance between kinase and phosphatase activity thus determines
whether or not $\sigma^F$ is released from its inactive complex with
SpoIIAB. SpoIIE accumulates on both sides of the asymmetrically
positioned septum and therefore has an increased activity in the
smaller compartment \cite{Arigoni95}. A quantitative model of the
regulatory network predicts that because of the low turn-over rate
most SpoIIE is bound by its substrate such that enzyme and substrate
increase together in the smaller compartment \cite{Iber_Spo_nature}.
According to the model, this combined increase is sufficient to
trigger the formation of micromolar concentrations of $\sigma^F$
holoenzyme in the prespore.

It is obvious from the above that the protein concentration ratio is
important. An excess of $\sigma^F$ or SpoIIAA compared to SpoIIAB
will result in free $\sigma^F$ and $\sigma^F$-dependent gene
expression while an excess of SpoIIAB will prevent SpoIIAA-dependent
$\sigma^F$ release. In the vegetative cell the sporulation proteins
are not detectable, and septation is preceded by $90-120'$ of gene
expression, dependent on the exact experimental conditions
\cite{Magnin97,Lucet99,Lord99}. Limiting the stochastic noise
inherent in protein expression can be expected to be crucial for
avoiding variations in the relative protein concentrations and the
resulting sporulation defects. Three of the four proteins in the
network are transcribed from genes in the \textit{spoIIA} operon
(Fig. \ref{Fig_operon_parallel}A). These genes are not only
co-transcribed into a single mRNA but are also  most likely to be
co-expressed since the translation of the three proteins appears to
be coupled, at least to some degree. This system therefore offers an
excellent opportunity to analyse the influence of transcriptional
and translational co-regulation of the sporulation genes on an
individual's survival, fitness.

Coupled translation is achieved when two genes are translated by the
same ribosome. Reinitiation of translation at a nearby start codon
after termination at the upstream gene is possible because ribosome
dissociation from the mRNA is a slow and energy-dependent process
\cite{McCarthy90}. There is currently no direct experimental
evidence for coupled translation of the \textit{spoIIA} operon. Such
coupling can, however, be postulated based on the arrangement of
genes \cite{Fort84}. The first two genes in the \textit{spoIIA}
operon (encoding SpoIIAA and SpoIIAB) overlap by four basepairs,
while the genes for SpoIIAB and $\sigma^F$ are interspaced by 11
basepairs (Fig. \ref{Fig_operon_parallel}A); coupled translation has
been documented for intercistronic distances of more than 60
basepairs \cite{McCarthy90}. The majority of genes that are
organized in operons are separated by distances comparable to those
found in the \textit{spoIIA} operon \cite{Salgado00}, so that the
studied system can be considered as representative of operons in
general. The efficiency of reinitiation depends on the distance as
well as the strength of the Shine-Dalgarno sequence
\cite{McCarthy90,Adhin90} which is, in general, located 5-13
basepairs upstream of a start codon and which binds to the
homologous 3' end of the 16S rRNA, a component of the 30S ribosomal
subunit. Moreover, the secondary structure of the mRNA can affect
lateral diffusion of the ribosomes \cite{Adhin90}.

According to the protein expression data for the \textit{spoIIA}
operon it appears that the last gene in the operon, $\sigma^F$, is
expressed at much lower levels than are SpoIIAA and SpoIIAB, while
SpoIIAB monomers may be expressed at equal or up to 3-times higher
levels compared to SpoIIAA \cite{Magnin97,Lucet99,Lord99}. The
weaker expression of a downstream gene (as is the case for
$\sigma^F$) can, in general, be accounted for by a weaker ribosomal
binding site which is removed far enough from the termination codon
of the upstream cistron that a considerable fraction of ribosomes
dissociate from the mRNA before translation can be reinitiated
\cite{McCarthy90}. It should be noted that while the transcriptional
and translational coupling will reduce the noise in the relative
SpoIIAB to $\sigma^F$ expression levels the unbinding of ribosomes
is necessarily a stochastic process and will therefore add a (low
level) of noise. The stronger expression of a downstream gene (as
may be the case for SpoIIAB relative to SpoIIAA) can, in general,
only be observed if a strong initiation sequence for the downstream
gene is occluded by mRNA secondary structure which is melted by the
ribosome that transcribes the upstream gene \cite{McCarthy90}. Such
a condition does not seem to be met by the gene for SpoIIAB, and
more accurate expression data will be necessary to establish whether
more SpoIIAB than SpoIIAA is expressed.

Available expression data can best be captured by an expression rate
for SpoIIAB dimers and SpoIIAA of $6 \times 10^{-9} M s^{-1}$ and of
$2 \times 10^{-9} M s^{-1}$ for $\sigma^F$ and SpoIIE
\cite{Iber_Spo_nature}; it should be noted that the simulation
yields qualitatively similar results if SpoIIAB monomers and SpoIIAA
are expressed at equal rates ($6 \times 10^{-9} M s^{-1}$), as long
as the $\sigma^F$ and SpoIIE expression rate is then reduced to
$10^{-9} M s^{-1}$ \cite{Iber_Spo_JMB}. As discussed in
\cite{Iber_Spo_JMB} the linear increase in the protein concentration
assumed here does not fully match the experimental observations.
There are, nonetheless,  two good reasons to use a linear model.
First of all, the data is too inaccurate and, in parts,
contradictory to be modeled exactly. Secondly, the chosen rates
correspond to the protein concentrations measured at the time of
septation \cite{Magnin97,Lucet99,Lord99}, the critical time point to
judge sporulation success. This is because in the cell the IIE
concentration increases more slowly than the other protein
concentrations and only increases sharply immediately before
septation \cite{Feucht02}. As a consequence, the greatest danger of
spontaneous uncompartmentalized activation of $\sigma^F$ is just
before septation, and this risk is fully assessed by the linear
expression model. Since our analysis focuses mainly at what happens
minutes before and after septation, individual fluctuations in the
global expression rates during the 2 hours preceding septation are
not important and the linear protein expression rates used should be
considered as an averaged protein expression rate per bacterium.

Our quantitative ordinary differential equation model is very
detailed - it comprises 50 dependent variables and 150 kinetic
constants to describe the dynamics of only four proteins; the reader
is referred to a detailed discussion of the model in the
Supplementary Material of \cite{Iber_Spo_nature}. Given its high
level of detail and accuracy the model predicts the phenotypes of
essentially all mutants for which the biochemical effect is known.
We can therefore expect that the predicted sporulation efficiencies
in response to changes in parameter values are realistic. In the
following we employ the model to quantify how far different levels
of stochastic noise in gene expression, as modulated by different
degrees of coupling of protein expression (that is by the coupling
of both transcription and translation), affect the sporulation
efficiency, that is the survival chances.

\section{Results and Discussion}
In the following we address how variations in the protein expression
rates affect the sporulation efficiency. Here we will look at the
effect of parallel changes in all protein expression rates as well
as at the effects of independent changes that skew the ratios of
protein concentrations. As the standard,``wild-type" protein
expression rates we use $6 \times 10^{-9} M s^{-1}$ for SpoIIAA and
SpoIIAB dimers and $2 \times 10^{-9} M s^{-1}$ for $\sigma^F$ and
SpoIIE \cite{Iber_Spo_nature}. After 120 minutes of protein
expression the septum forms and SpoIIE accumulates on both sides of
this septum. This is modeled by a four-fold increase in the
concentration of SpoIIE, together with its associated substrate
(phosphorylated SpoIIAA) in the prespore. As before we define a
successful sporulation event by the requirement that before
septation the concentration of $\sigma^F \cdot$ RNA polymerase
holoenzyme does not exceed 0.4 $\mu$M while after septation the
concentration exceeds one micromolar \cite{Iber_Spo_nature}.

If the protein expression rates are all varied in parallel, that is
by a common factor as denoted on the horizontal axis in Figure
\ref{Fig_operon_parallel}B, we find that the predicted sporulation
efficiency is not affected as long as a minimal expression rate is
kept to provide sufficient $\sigma^F$ for binding to the RNA
polymerase (Fig. \ref{Fig_operon_parallel}B, grey lines). If the
expression of SpoIIE is kept constant (in order to reflect that this
protein is transcribed from a different locus and may therefore vary
independently) then an independent 2.5-fold increase in the other
sporulation proteins can still be tolerated before the relative
activity of the phosphatase becomes too weak (Fig.
\ref{Fig_operon_parallel}B, black lines). An even higher independent
increase in the expression of the \textit{spoIIA} genes can be
tolerated if we assume that the expression of the \textit{spoIIA}
and \textit{spoIIE} genes is at least weakly correlated such that a
large increase in the expression of the \textit{spoIIA} genes is
accompanied by a small increase in the expression of the
\textit{spoIIE} genes (Fig. \ref{Fig_operon_parallel}C). Such a
correlation is not unexpected considering that variations in gene
expression are the result of both intrinsic and extrinsic noise. The
latter, which reflects cell-to-cell variation in the concentration
of other molecular species such as the RNA polymerase, will affect
all genes similarly. We can conclude that the independent regulation
of the \textit{spoIIA} and \textit{spoIIE} genes is unlikely to
generate a major risk of failed sporulation. Separation of the
\textit{spoIIA} and \textit{spoIIE} genes on the bacterial
chromosome, on the other hand, has benefits because it ensures that,
upon septation, each compartment retains one copy of \textit{spoIIE}
while initially (for the first $10-15'$) two copies of
\textit{spoIIA} are in the mother cell but none in the prespore
\cite{Frandsen99}. This initial transient genetic imbalance may
protect the mother cell from a relative increase of \textit{spoIIE}
to \textit{spoIIA} gene products \cite{Iber_Spo_JMB}.

If the expression levels of the genes in the \textit{spoIIA} operon
are varied independently of each other, the tolerance of the network
to variations in gene expression drops substantially. In particular,
if SpoIIAB and SpoIIAA are no longer co-regulated, the network is
sensitive to rather small changes (Fig. \ref{Fig_operon_parallel}D,
grey lines and circles). Thus if the SpoIIAA expression rate remains
fixed and the SpoIIAB expression rate increases by $60\%$
(corresponding to the factor 1.6 on the horizontal axis in Fig.
\ref{Fig_operon_parallel}D), then sporulation is predicted to fail;
$60\%$ variation from the mean is a noise level observed in
bacterial (\textit{E. coli}) expression systems \cite{Elowitz02}. On
the other hand, if SpoIIAA and SpoIIAB remain co-regulated but
$\sigma^F$ expression is regulated independently (Fig.
\ref{Fig_operon_parallel}D, black lines), the network is rather
robust to variations in gene expression as long as the expression of
SpoIIAB is increased more than the expression of $\sigma^F$ and the
overall $\sigma^F$ concentration remains high enough to form
micromolar concentrations of the holoenzyme. The transcriptional
coupling together with a strong translational coupling of SpoIIAA
and SpoIIAB therefore substantially increases the robustness of the
network to fluctuations in gene expression. Stochastic variations in
the relative rate of $\sigma^F$ translation, on the other hand, is
not as detrimental as long as the translation efficiency for
$\sigma^F$ is lower than for SpoIIAA and SpoIIAB, as can be achieved
by a weaker ribosomal binding site and the resulting (stochastic)
dissociation of ribosomes. An advantage of preferential dissociation
of the ribosomes before translating the gene for $\sigma^F$ is that
the bacterium saves the energy that would otherwise be required to
translate, and subsequently degrade, unnecessary (harmful) copies of
$\sigma^F$. Considering that $\sigma^F$ comprises 255 amino acids
and linkage of each amino acid requires the equivalent of 4 ATPs the
energy by not translating and degrading 10 $\mu$M $\sigma^F$
corresponds to more than 10 mM ATP, which is a considerable amount
considering that the bacterial ATP concentration is 1-3 mM
\cite{Jolliffe81,Guffanti87,Hecker88} and sporulation is a response
to starvation, that is energy deprivation.

In a last step we can now quantify the impact of gene organisation
on sporulation efficiency, and therefore fitness. For this we assume
that the gene expression levels in the cell population follow a
normal distribution with variance $\eta$ around the mean value.
Given the complex regulation pattern of gene expression, gene
expression levels are unlikely to be distributed exactly normally. A
normal distribution is, however, still likely to provide an
approximation no worse than what could be obtained with a detailed
model of the regulatory process in the absence of sufficient data to
determine all required parameter values \cite{Swain04}. Sporulation
efficiency is determined as the fraction of simulation runs for
which the concentration of $\sigma^F \cdot$ RNA polymerase
holoenzyme does not exceed 0.4 $\mu$M before septation and exceeds
one micromolar after septation \cite{Iber_Spo_nature}. For each
condition the mean sporulation efficiency and standard deviation are
calculated from 100 independent runs that are carried out 10 times.
In each run the protein expression rates were set randomly such that
overall the respective distributions of the protein expression rates
were obtained. Determination of the sporulation efficiency for $\eta
\in [0, 1]$ shows that as long as the sporulation genes are
translationally coupled, even high variances hardly affect the
sporulation efficiency (Fig. \ref{Fig_operon_stoch}A, black lines).
The sporulation efficiency is even higher at high noise level,
$\eta$, if \textit{spoIIE} expression co-varies with \textit{spoIIA}
expression, at least weakly (Fig. \ref{Fig_operon_stoch}B). A
lengthening of the transcription time, (that is a delay in
septation) when transcription levels are too low to generate
sufficient $\sigma^F$ until septation will further increase
robustness to fluctuations in the rate of protein expression. Such a
dependency of the time point of septation on the protein (and in
particular the SpoIIE) concentration is in agreement with
experiments \cite{Khvorova98,Ben-Yehuda02} and might explain the
large variance in the delay between the onset of sporulation and
septation that is observed under different sporulation conditions.
Low levels of additional stochastic noise in $\sigma^F$ expression
(broken lines), as may arise because of the stochastic dissociation
of ribosomes, also has rather little impact and confirms that the
weak coupling of SpoIIAB and $\sigma^F$ translation does not
substantially reduce sporulation efficiency. If on the other hand
\textit{spoIIAB} is removed from the operon and controlled
independently by the same promotor then the sporulation efficiency
drops rapidly (Fig. \ref{Fig_operon_stoch}A, blue lines). This is in
good quantitative agreement with experiments which find that the
sporulation efficiency drops to $40-80 \%$ of wildtype levels
\cite{Dworkin01}, especially when considering that $\eta \sim [0.3,
0.6]$ for these expression levels \cite{Elowitz02}. If
\textit{spoIIAA} is moved instead, then the effect is reduced
(Joanna Clarkson, personal communication) as also predicted by the
model (Fig. \ref{Fig_operon_stoch}A, grey lines).

It should be noted that this drop in sporulation efficiency has
previously been accounted for by the loss of the transient genetic
imbalance when \textit{spoIIAB} is moved to a chromosomal position
close to the origin of replication \cite{Dworkin01}. The transient
lack of SpoIIAB expression in the prespore together with accelerated
degradation of unbound SpoIIAB \cite{Pan01} had been suggested to
enable $\sigma^F$ release \cite{Dworkin01}. However, we have shown
previously that the transient genetic imbalance does not affect
$\sigma^F$ release on the timescale on which it persists
\cite{Iber_Spo_JMB}, and stochastic effects are therefore a much
more likely explanation for the observed phenotype of the mutants.

We conclude from the analysis of this well studied model system that
the protection from stochastic variation in the expression rate of
interacting proteins can substantially increase viability, and
therefore constitutes a driving force for gene clustering and
co-regulation. Whilst the importance of gene dosage had been
recognized before \cite{Veitia02}, and underexpression and
overexpression of protein complex subunits in yeast had been shown
to lower fitness \cite{Papp03}, this study reveals that much smaller
variances, as can result from stochastic effects, can already have
substantial detrimental effects. The detailed analysis of the
expression of the sporulation proteins therefore demonstrates the
optimized character of gene regulation and suggests that
co-regulation of genes serves to optimize cellular network dynamics
in spite of the inherent noise in all biological processes.

\paragraph{Acknowledgements}
I wish to thank Iain D. Campbell, Joanna Clarkson, and Michael D.
Yudkin for many valuable discussions, and Iain D. Campbell for his
critical reading of the manuscript. The work was supported by a DTA
EPSRC studentship as well as by a Junior Research fellowship held at
St John's College, University of Oxford.

\bibliography{C:/Dagmar/Science/Texfiles/refs}
\bibliographystyle{C:/Dagmar/Science/Texfiles/bst/Dagmar/embo/embo}

\newpage

\paragraph{Figures}
\paragraph{Figure 1: An overview of the interactions in the network that
controls $\sigma^F$ activity in \textit{Bacillus subtilis}.} For
details see text. The figure is a reproduction of Figure 1 in
\cite{Iber_Spo_nature}.

\paragraph{Figure 2: The impact of parallel and random variations in the
expression of \textit{spoIIE} and \textit{spoIIA} genes on
$\sigma^F$ release.} \textbf{(A)} The \textit{spoIIA} operon
comprises the genes for SpoIIAA, SpoIIAB, and $\sigma^F$. The genes
for SpoIIAA and SpoIIAB overlap; the genes for SpoIIAB and
$\sigma^F$ are separated by 11 basepairs. \textbf{(B)} The
regulatory network is robust to parallel variations in gene
expression. The predicted concentration of $\sigma^F
\cdot$RNApolymerase holoenzyme before (dashed lines) and after
septum formation (continuous lines) if either all (grey lines) or
all protein expression rates except for the one of SpoIIE (black
lines) were increased by the factor on the horizontal axis compared
to the standard reference rates ($6 \times 10^{-9} M s^{-1}$ for
SpoIIAA and SpoIIAB dimers and $2 \times 10^{-9} M s^{-1}$ for
$\sigma^F$ and SpoIIE \cite{Iber_Spo_nature}). \textbf{(C,D)} The
expression rate combinations for which septation-dependent
$\sigma^F$ release is possible (between the lines) or not (outside
the area marked by lines). \textbf{(C)} The impact of differential
regulation of \textit{spoIIE} and \textit{spoIIA} expression. The
vertical and horizontal axes indicate the fold variation in the
\textit{spoIIE} and \textit{spoIIA} expression rates respectively,
compared to the standard reference rates. \textbf{(D)} The impact of
differential regulation of the expression of genes encoded in the
\textit{spoIIA} operon. The vertical axis indicates the fold
variation in the expression of SpoIIAA (circles), $\sigma^F$ (black
lines), or SpoIIAA and $\sigma^F$ (grey lines). The horizontal axis
indicates the fold variation in the expression of SpoIIAB and of any
other protein whose expression is coupled to the one of SpoIIAB
(which are those genes in the \textit{spoIIA} operon not reported on
the vertical axis). The sudden jump observed at a high SpoIIAB to
$\sigma^F$ ratio (lower black line) is the consequence of impaired
$\sigma^F$ release when the relative SpoIIAB concentration is too
high.

\paragraph{Figure 3: The impact of stochastic variation in gene expression
on sporulation efficiency.} \textbf{(A)} The fraction of successful
sporulation events (as defined in the text) dependent on the
variance in gene expression if expression of the \textit{spoIIA}
genes is either coupled (black lines), the expression of SpoIIAB and
$\sigma^F$ is coupled (grey lines), or the expression of SpoIIAA and
$\sigma^F$ is coupled (blue lines). SpoIIE is expressed throughout
at the standard rate of $2 \times 10^{-9}$ M$^{-1}$ s$^{-1}$. The
broken lines show the effect of an additional independent normal
variation in the rate of $\sigma^F$ expression with $\eta_S = 0.1$
(dashed lines) or $\eta_S = 0.3$ (dotted lines) from the coupled
rates. If $\sigma^F$ is one of the coupled rates then $\sigma^F$
expression is varied both together with its coupling partner and
additionally independently to reflect the additive levels of noise
acting at the initiation of translation and the
re-initiation/dissociation step.  \textbf{(B)} The fraction of
successful sporulation events (as defined in the text) dependent on
the variance in gene expression if expression of the \textit{spoIIA}
and \textit{spoIIE} genes is coupled (to assess the benefits of
correlated expression), and an additional noise term $\eta_E$ is
added to the expression of \textit{spoIIE} with $\eta_E =0.1$ (black
continuous line), $\eta_E =0.3$ (dotted line), or $\eta_E =0.6$
(dashed line);  $\eta_E$ assesses the effects of independent
promotors and spatial heterogeneity in the concentration of
transcription and translation factors. The red line is identical to
the continuous black line in panel A (noise in coupled
\textit{spoIIA} expression, SpoIIE expressed at $2 \times 10^{-9}$
M$^{-1}$ s$^{-1}$). Mean and standard deviation are based on 10
times 100 independent runs.

\newpage

\begin{figure} [!t]
\begin{center}
\includegraphics[width=0.5\textwidth]{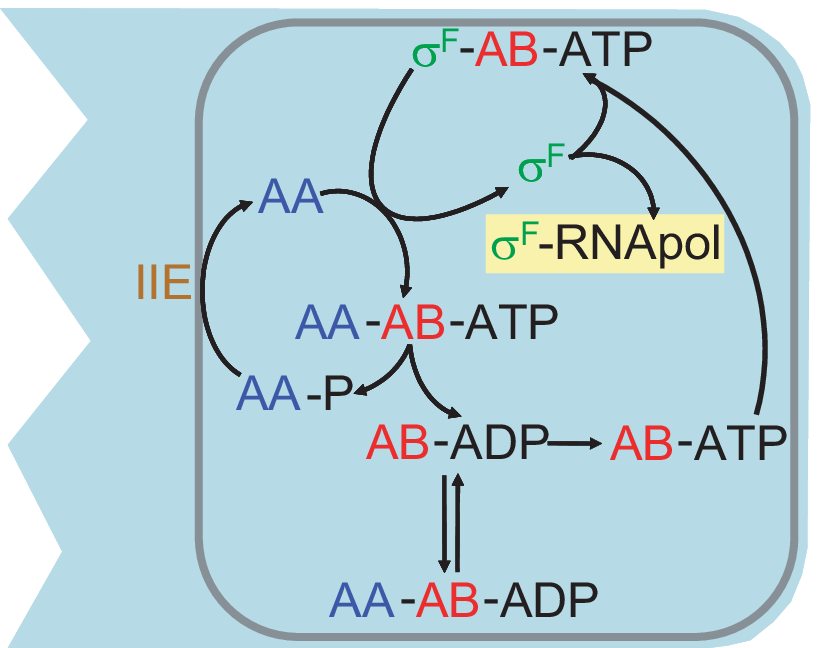}
\caption []{\label{Fig_model} }
\end{center}
\end{figure}

\begin{figure} [!t]
\begin{center}
\includegraphics[width=0.5\textwidth]{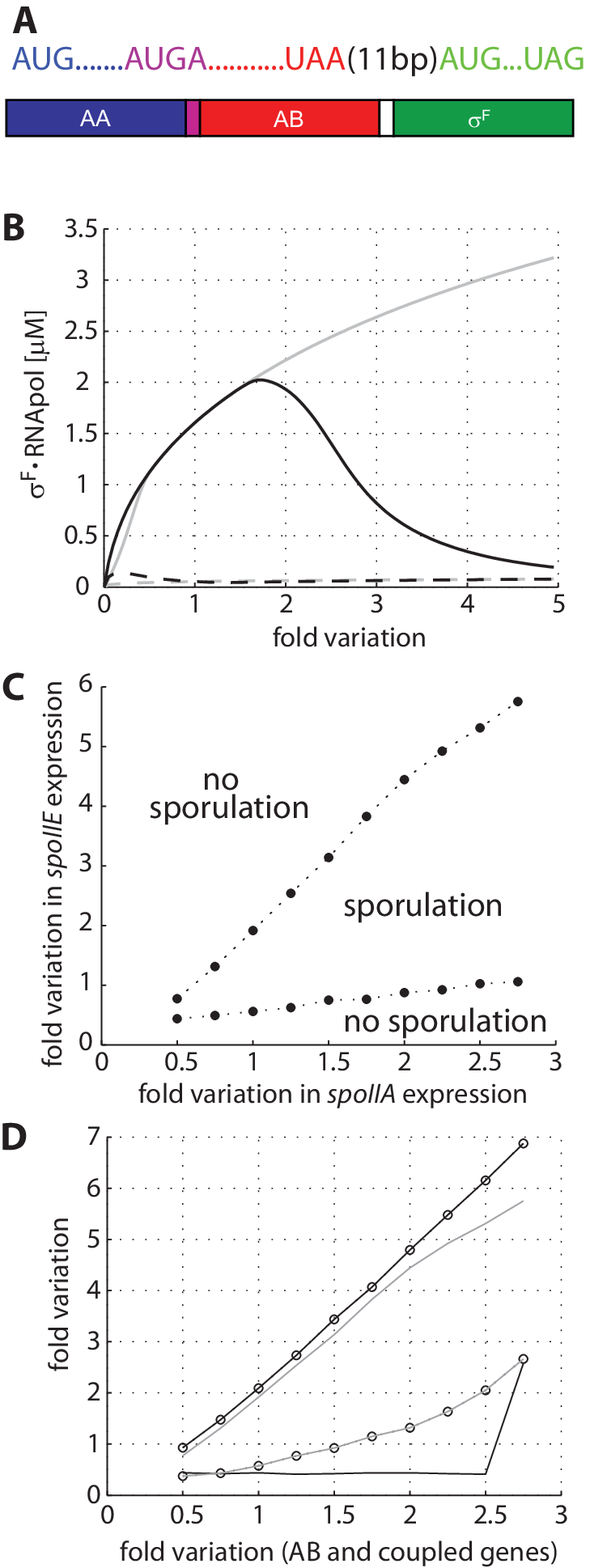}
\caption []{\label{Fig_operon_parallel} }
\end{center}
\end{figure}

\begin{figure} [!t]
\begin{center}
\includegraphics[width=0.5\textwidth]{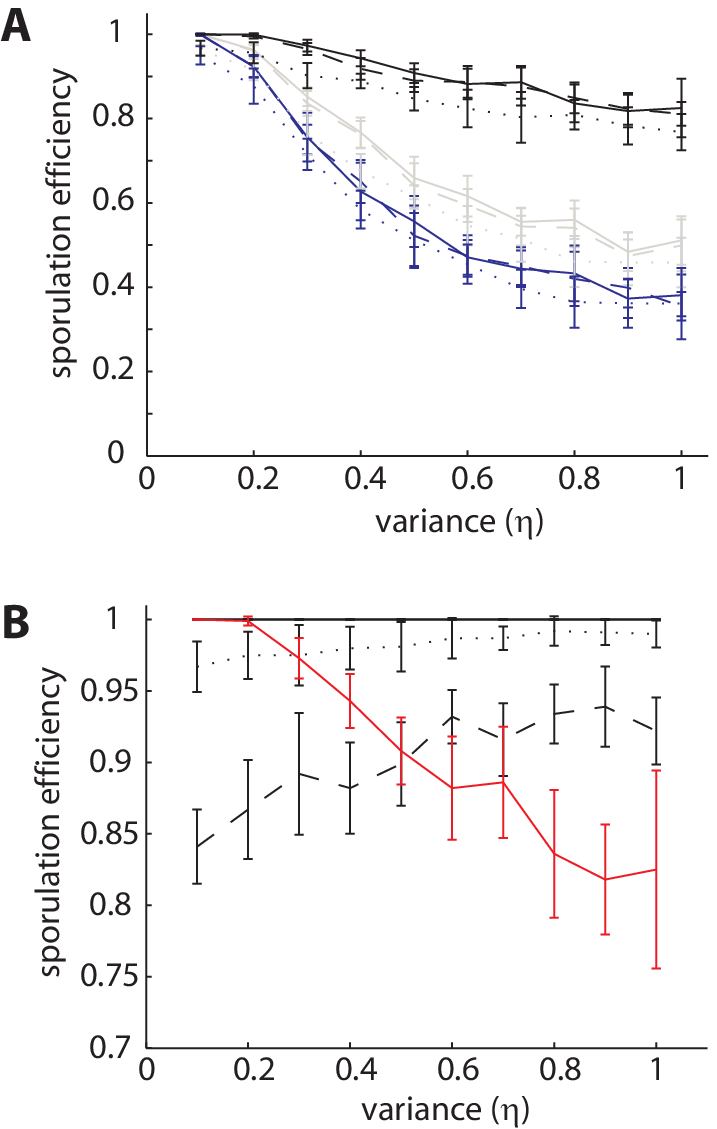}
\caption []{\label{Fig_operon_stoch} }
\end{center}
\end{figure}

%\begin{figure} [!t]
%\begin{center}
%\includegraphics[width=0.9\textwidth]{Fig_robustness_operon.eps}
%\caption []{\label{Fig_SpoIIAB degrad} }
%\end{center}
%\end{figure}

\end{document}